\documentclass[aps, prl, amsmath, showpacs, preprintnumbers, superscriptaddress, twocolumn, amssymb]{revtex4-1}
\usepackage{graphicx}
\usepackage{mathptmx, textcomp}
\usepackage[latin1]{inputenc}

\begin{document}
\title{Dynamics of a cold trapped ion in a Bose-Einstein condensate}
\author{Stefan Schmid}
\author{Arne H\"arter}
\author{Johannes Hecker Denschlag}
\affiliation{Institut f\"ur Quantenmaterie, Universit\"at
 Ulm,
 89069 Ulm, Germany}
\affiliation{Institut f\"ur Experimentalphysik und Zentrum f\"ur Quantenphysik, Universit\"at
 Innsbruck,
 6020 Innsbruck, Austria}

\date{\today}


\begin{abstract}
We investigate the interaction of a laser-cooled trapped ion (Ba$^+$ or Rb$^+$) with an optically confined $^{87}$Rb Bose-Einstein condensate (BEC). The system features interesting dynamics of the ion and the atom cloud as determined by their collisions and their motion in their respective traps. Elastic as well as inelastic processes are observed and their respective cross sections are determined. We demonstrate that a single ion can be used to probe the density profile of an ultracold atom cloud.
\end{abstract}

\maketitle

The realization of a charged quantum gas, formed by laser-cooled, trapped ions and ultracold neutral atoms, offers intriguing perspectives for  a variety of novel experiments. In addition to studying atom-ion collisions and chemical reactions in a regime where only one or few partial waves contribute, interesting phenomena such as charge transport in the ultracold domain \cite{Cot2000a}, polaron-type physics \cite{Mas2005,Kal2006,Cuc2006}, and novel atom-ion bound states \cite{Cot2002} can be investigated. Further, the production of cold, charged molecules in a well-defined quantum state is an important goal in molecular physics \cite{Sta2010,Sch2010a}.
Techniques and tools that are widely used to control neutral atomic quantum gases, e.g.~Feshbach resonances \cite{Chi2010}, may be employed in atom-ion collisions as well \cite{Idz2009}. First observations of cold collisions with trapped atoms and ions in the mK regime have been made using Yb$^+$ ions
and a magneto-optical trap for Yb \cite{Cet2007,Gri2009} and very
recently with a Bose-Einstein condensate (Yb$^+$, Rb)
\cite{Zip2010}. Here we present our investigations where we use a
defined small number of Ba$^+$ or Rb$^+$ ions which are immersed
in an ultracold cloud of Rb atoms, which is either Bose-Einstein
condensed or at sub-\textmu K temperature. We find that at our
current settings micromotion of the trapped ions plays an important
role in our experiments. The elastic cross sections are large as
expected. Inelastic processes are in general strongly suppressed,
which is important for planned future experiments. Charge transfer
is directly identified in our experiments and found to be the
dominant inelastic collision channel. As a first application we
show how a trapped ion can be used to locally probe the density
profile of a condensate.

In the inhomogeneous electrical field of an ion, a neutral atom is
polarized and attracted towards the ion. This interaction can be
expressed as a long-range polarization potential $V(r) = - C_4/2r^4$
with $C_4 = q^2 \alpha /4\pi\epsilon_0$, where $q$ is the
charge of the ion and $\alpha$ the dc polarizability of the atom.
The characteristic radius of the potential is given by $r^* =
\sqrt{\mu C_4 / \hbar}$, where $\mu$ is the reduced mass. For a
$^{87}$Rb atom ($\alpha= 4.7\times 10^{-31}\,$m$^3$
\cite{Tea1971}) interacting with a $^{138}$Ba$^+$ ion, this
characteristic radius is $r^*= 295\,$nm. This is much larger than
the typical length scale of the neutral atom-atom interaction
potential as given by the van der Waals radius, which for
$^{87}$Rb is $R_{vdW}\approx 4\,$nm \cite{Chi2010}. For collision
energies $E$ above $ k_\textrm{B} \times 100\,$\textmu K, the cross section for
elastic scattering can be approximated by the semiclassical
estimate $\sigma_{\textrm{el}} = \pi(\mu
C_4/\hbar^2)^{1/3}(1+\pi^2/16) E^{-1/3}$ \cite{Cot2000b}. At lower energies a full quantum mechanical treatment is necessary, which takes into account individual partial waves. In addition to elastic collisions, inelastic processes can also occur, such as charge transfer Rb$\,$+$\,$Ba$^+\rightarrow\,$Rb$^+$+Ba$\,$ or molecule formation Rb$\,$+Rb$\,$+$\,$Ba$^+\rightarrow\,$Rb$\,$+$\,$(BaRb)$^+$.

\begin{figure}
 \includegraphics[width=8.0cm] {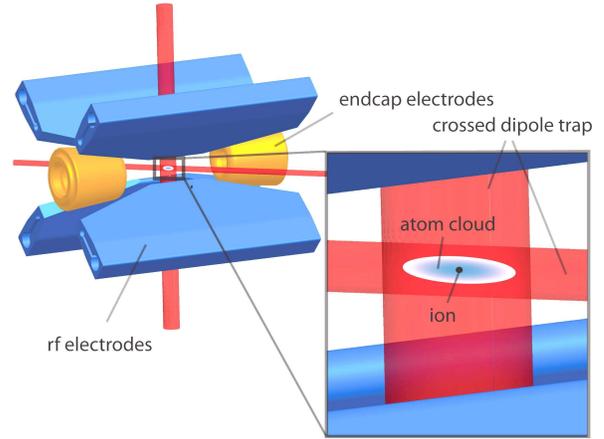}
 \caption{Combined atom-ion trap: A linear radiofrequency Paul trap is used to store ions \cite{Win1998}, whereas the ultracold atoms are confined in a crossed optical dipole trap. By precisely overlapping the positions of these two traps, a single ion can be immersed into the center of the ultracold neutral atom cloud.}
 \label{fig1}
\end{figure}

Here we present the first results obtained with our novel hybrid apparatus,
where both species (Ba$^+$, Rb) are first trapped and cooled in separate parts
of a UHV chamber in order to minimize mutual disturbance and are then brought together. We use a linear Paul trap into which we can load a well-defined number of
$^{138}$Ba$^{+}$ ions (typically 1) by photoionizing neutral Ba atoms from a
getter source. Photoionization is done with a diode laser at
413$\,$nm, which drives a resonant two-photon transition from the
ground state to the continuum via the $^3$D$_1$ state. The $^{138}$Ba$^{+}$ ion
which has no hyperfine structure is Doppler-cooled to mK
temperatures using a laser at 493$\,$nm. Another laser at
650$\,$nm repumps from a metastable D$_{3/2}$ state. The ion's
fluorescence is collected with a lens of NA$\,$=$\,$0.2 and then
detected using an electron multiplying charged-coupled device
(EMCDD) camera. The linear Paul trap (see Fig.~1) consists of four
blade electrodes, which are placed at a distance of 2$\,$mm from
the trap axis, and two endcap electrodes
 at a distance of 7$\,$mm from the trap center. It is operated
 with a radiofrequency (rf) drive of $\Omega =
2\pi\times$5.24$\,$MHz with an amplitude of
1400$\,$V$_\textrm{pp}$ and a dc endcap voltage of about 100$\,$V.
 With these parameters we measure the axial and the radial trap frequencies to be
$\omega_{\textrm{\,I,ax}} \approx 2\pi\times$80$\,$kHz and
$\omega_{\textrm{\,I,rad}} \approx 2\pi\times$200$\,$kHz,
respectively.

The sample of ultracold Rb atoms is prepared similarly as
described in \cite{Tha2005}. Atoms from a magneto-optical trap are
magnetically transferred over a distance of 43$\,$cm into a
quadrupole Ioffe configuration (QUIC) trap. Evaporative cooling
yields a cloud of 3$\times$10$^6$ $^{87}$Rb atoms in state
\linebreak $|\,F=1,\;m_F=-1\rangle$ at a temperature of about 1$\,$\textmu K.
From the magnetic trap we load the atom cloud into a vertical 1-dimensional optical lattice trap formed by two counterpropagating laser beams  (diameter $\approx$ 500$\,$\textmu m) at a wavelength of 1064$\,$nm with a total power of 2$\,$W. We use the lattice as an elevator to transport the ultracold atoms upwards into the center of the Paul trap which is located 30$\,$cm above the QUIC trap. The lattice is moved by changing the relative detuning of the two beams in a controlled manner \cite{Sch2006}. The transport efficiency reaches more than 60$\%$.
During transport the depth of the lattice is sufficiently small ($\approx k_\textrm{B} \times 10\,$\textmu K)
so that evaporation keeps the atom temperature at about 1$\,$\textmu K.
After transport  the atomic sample is loaded into a crossed optical dipole trap ($\lambda$ = 1064$\,$nm).
By lowering the trap depth, we evaporatively cool the atoms and end up with a BEC of about 10$^5$ atoms
confined in a dipole trap with trap frequencies of about $\omega_\textrm{Rb}^\textrm{(x,y,z)}=(60\,$Hz, 60$\,$Hz, 8$\,$Hz). The condensate is detected via standard absorption imaging.

The BEC is initially located about 300$\,$\textmu m away from the ion. We move the ion within 2$\,$ms along the Paul trap axis into the BEC by changing the endcap voltages. The cooling lasers for the ion are switched off in order to ensure that the ion relaxes into its electronic ground state \linebreak $|F=1/2,\;m_F=\pm 1/2\rangle$ and to avoid changes in the atom-ion collision dynamics due to the cooling radiation.

\begin{figure}
    \includegraphics[width=8.0cm] {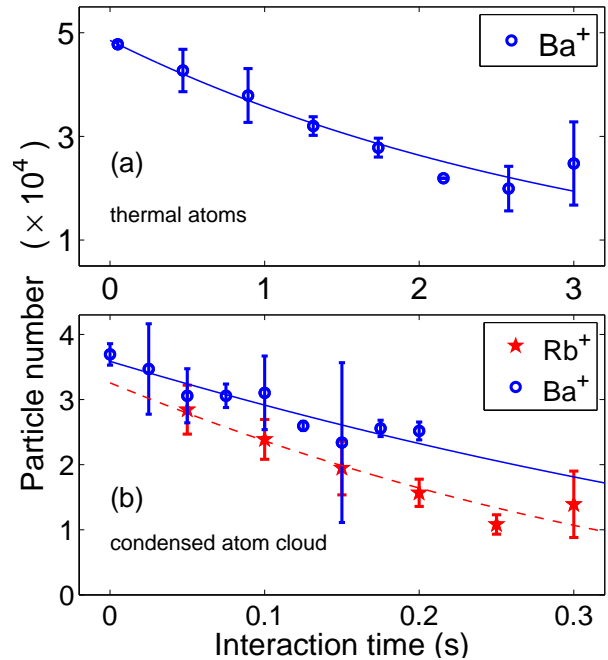}
    \caption{Number of remaining Rb atoms as a function of time as a sample of Rb atoms interacts with an ion. \textbf{(a)} A single Ba$^+$ ion is immersed into the center of a thermal Rb cloud. The line is an exponential decay fit. \textbf{(b)} A single ion (Ba$^+$ or Rb$^+$) is immersed in a Rb Bose-Einstein condensate. The lines are fits based on our simple model described in the text.}
    \label{fig2}
\end{figure}

Initially the temperature of the ion is on the order of a few mK, which is much larger than the optical trap
depth of about $k_B\times$1$\,$\textmu K. Thus almost any collision with a cold atom leads to the atom being lost from the trap.

As a consequence, one could naively expect, that the atom loss stops
after a few collisions, once the ion is sympathetically cooled to
atomic temperatures. However, when we measure the number of Rb
atoms remaining in the trap as a function of the interaction time,
we find a continuous loss of atoms (Fig.~2). It is the driven micromotion of the rf trap, which
\cite{Ber1998} is responsible for this continuing loss.
In an atom-ion collision, energy can be redistributed among all motional degrees of
freedom, enabling also the flux of micromotion energy to secular
motion. After each collision, micromotion is quickly restored by
the driving rf field. An equilibrium between the energy that is
inserted by the driving field and the energy taken away by the
lost atoms is reached. Thus the minimal temperature of a sympathetically-cooled ion stored in a rf trap is determined by the amount of micromotion of the ion \cite{Maj1968}. Stray electric fields
which shift the ion position away from the zero of the rf
potential increase micromotion, leading to the so-called excess
micromotion \cite{Ber1998}. By applying dc electric fields along the radial direction this part of the micromotion is tunable.

Fig.~2 shows the elastic collision measurements with either (a) a thermal cloud of atoms (temperature
$T_\textrm{Rb}\,$=$\,$80$\,$nK, which is just above $T_c$) or (b) a BEC. In
the following we analyze these data in detail starting out with
the case for thermal atoms. In general, we have checked that
losses due to collisions with the background gas are negligible.
The loss of atoms is then described by $\dot{N} = -\tilde{n}
\sigma_\textrm{el} v_\textrm{I}$, where $\tilde{n}$ is the density
of the atom cloud at the position of the ion, $\sigma_\textrm{el}$
the elastic atom-ion cross section and $v_\textrm{I}$ the velocity
of the ion. In principal all three quantities are unknown. Based on our measurements and additional constraints of the following model we can still give estimates for them.

\begin{figure}
 \includegraphics[width=8.0cm] {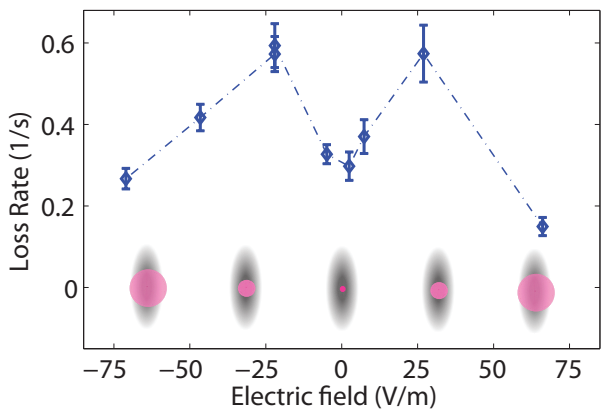}
 \caption{Loss rate of a thermal atom cloud when a static electric field is applied.
 In the presence of a field, the energy of the Ba$^+$ ion and thus the amplitude
 of its secular motion are increased. As a result the sphere of depletion becomes
 larger, leading to an enhanced atom loss rate. For fields
 $\mid \mathcal{E}_{\textrm{dc}}\mid > 30\,$V/m, the amplitude of the secular
 motion is so large, that the ion spends a significant amount of time in a region
 of lower atom density. Therefore the loss rate decreases at high fields.}
 \label{fig3}
\end{figure}

It is important to note that the density profile of the atom
cloud is depleted locally due to the collisions with the well-localized ion. According to
its mass $m_\textrm{I}$, energy $E_\textrm{I} = m_\textrm{I} v_\textrm{I}^2/2$, and trap frequency
$\omega_{\textrm{\,I}}$ the ion will be localized inside a region (which for simplicity we choose to be spherical) of radius $r_0 = \sqrt{E_\textrm{I}/m_\textrm{I} \omega_\textrm{\,I}^2}$. Inside this sphere we assume a homogeneous
density $\tilde{n}$. Right outside the sphere, the density is given by $n=N\bar{\omega}_\textrm{Rb}^3(m_\textrm{Rb}/2\pi k_B T_\textrm{Rb})^{3/2}$, which is the peak density of a thermal cloud of Rb atoms (mass $m_\textrm{Rb}$) confined in a harmonic trap with mean trap frequency $\bar{\omega}_\textrm{Rb}=(\omega_\textrm{Rb}^\textrm{(x)}\omega_\textrm{Rb}^\textrm{(y)}\omega_\textrm{Rb}^\textrm{(z)})^{1/3}$. The measured atom loss $\dot{N}$, which at $t = 0$ is $\dot{N} = 1.5\times 10^4\,$s$^{-1}$ (as can be read off from Fig.~2), must match the net flux of atoms into the ``sphere of depletion", leading to the balance condition $\dot{N} = \pi r_0^2 (n-\tilde{n}) v_\textrm{th}$. Here the thermal velocity of the atoms is given by $v_\textrm{th} = \sqrt{k_B T_\textrm{Rb} /8\pi m_\textrm{Rb}}$. By setting $\tilde{n}=0$ in the balance condition we get a lower bound for the ion energy, which for our parameters is equal to $k_B\times$17$\,$mK. We can also estimate an upper limit of the ion energy from our method of compensating micromotion. The compensation is based on minimizing the position shift of the ion when the rf amplitude is changed. With this method we can reduce the DC electric field at the position of the ion to below 4$\,$V/m, corresponding to a maximum ion energy of
$k_B\times$40$\,$mK \cite{Ber1998}. By taking the midpoint between the two bounds we estimate the ion
energy to be about $k_B\times$30$\,$mK. Plugging this energy into the expressions above we get $r_0 \approx 1.45\,$\textmu m and a density of $\tilde{n}
\approx 0.45\,n$. Moreover we can determine the cross section
$\sigma_\textrm{el}$ to be $1.9 \times 10^{-14}\,$m$^2$. This value is in rough agreement with the semiclassical estimate $\sigma_\textrm{el}^\textrm{semi} = 9 \times 10^{-15}\,$m$^2$.

A slightly different analysis can be done for the
measurement with the BEC shown in Fig.~2(b). Again, collisions of the atoms with the
trapped ion will lead to a sphere of depletion within the
condensate. However, the flux of atoms into the sphere is now
driven by mean field pressure rather than thermal motion. From
numerical calculations using a 3D Gross-Pitaevskii equation with
an absorptive (imaginary) potential term, we find that the flux
into the sphere can be approximated by $4\pi
r_0^2\,\tilde{n}\,v_e$, as long as $\tilde{n} \gtrsim 0.1\,n$. Here
$v_e = 4\hbar\sqrt{2\pi a (n-\tilde{n})}/m_\textrm{Rb}$ is the
velocity with which the atoms enter the sphere, $a = 5.61\,$nm is the
Rb-Rb s-wave scattering length and $n = \left(15m_\textrm{Rb}^3\bar{\omega}_\textrm{Rb}^3 N/\hbar a^{3/2}\right)^{2/5}/8\pi$ is the peak density of the condensate. By equating the net flux to the
measured lossrate of the BEC at $t = 0$, $\dot{N} = 8\times
10^4\,$s$^{-1}$ (see Fig.~2), we find that the ion energy has to
be larger than $k_\textrm{B}\times$1$\,$mK.
Our analysis shows, that on the other hand the ion energy has to be smaller than in the thermal case. Indeed, to minimize the temperature of the trapped ion, a comparatively large effort was made to compensate micromotion for the experiments with the BEC. Given that $\sigma_\textrm{el}/\sigma_\textrm{el}^\textrm{semi} \approx 2$ as in the thermal case, our data suggests a reasonable ion energy of about $k_\textrm{B}\times$5$\,$mK. This implies $r_0 \approx 0.8\,$\textmu m, a density of $\tilde{n} \approx 0.1\,n$ and a cross section of
$\sigma_\textrm{el} \approx 3.1 \times 10^{-14}\,$m$^2$. We also measured the loss rate for a single Rb$^+$ ion and found it to be about the same as with Ba$^+$ (Fig.~2). This is not surprising, since the atom loss rate depends only weakly on the ionic mass and the inner structure of the ion is irrelevant in the semiclassical regime.

\begin{figure}
 \includegraphics[width=8.0cm] {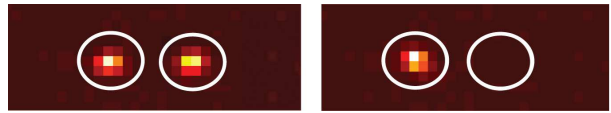}
 \caption{\textit{Left}: Fluorescence image of two Ba$^+$ ions. \textit{Right}: Fluorescence of one $^{138}$Ba$^+$ next to an unknown dark ion . We infer the existence of the dark ion from the position of the Ba$^+$.}
 \label{fig5}
\end{figure}

Following this discussion, we expect the net flux into the sphere and
thus the atom loss rate to rise when we intentionally increase the
dc electric field. For small fields this is also what we observe
in the experiment (Fig.~3). The maximum loss rate is observed at
an electric field of $\mathcal{E}_{\textrm{dc}} \approx 30 \,$V/m,
where the amplitude of the ion's secular motion is $r_0 \approx
20\,$\textmu m. This value is comparable to the size of the atom cloud,
which has an extension of about $15\,$\textmu m along the radial and
$80\,$\textmu m along the axial direction. The model of a well-localized
ion and a sphere of depletion is clearly no longer valid in
this regime. We explain the decrease of the atom loss rate for
even higher fields by the fact that the ion spends a significant
amount of time in regions of lower atom density.

In addition to the elastic processes discussed so far, we have also investigated inelastic atom-ion collisions. For this we load two $^{138}$Ba$^+$ ions into the ion trap. Typically after a time corresponding to $10^4$-$10^5$ elastic atom-ion collisions, the fluorescence of one of the Ba$^+$ ions is lost (Fig.~4). Since the position of the remaining bright
Ba$^+$ ion does not change, we infer that the other Ba$^+$ ion has been replaced by an unknown dark ion formed in a reaction. We can determine the mass of the dark ion by measuring the radial center-of-mass (COM) trap frequency. For this we modulate the amplitude of the rf voltage. When the modulation frequency is equal to the trap frequency the ion motion is resonantly excited and a drop of the Ba$^+$ fluorescence signal is observed. We only observe two resonance frequencies at 245$\,$kHz and 400$\,$kHz. We have checked that Ba$^+$ gives rise to the 245$\,$kHz resonance. Using the scaling properties of the analytic expression for the COM mode \cite{Win1998}, the 400$\,$kHz resonance then corresponds to the mass of $^{87}$Rb$^+$. We conclude that the dominant inelastic process in our system is the charge transfer process Rb$\,$+$\,$Ba$^+\rightarrow\,$Rb$^+$+Ba$\,$ with a cross section $\sigma_\textrm{ch.ex.}$ ranging between $10^{-19}\,$m$^2$ and $10^{-18}\,$m$^2$. Our charge transfer results are comparable to the ones observed for the heteronuclear case of (Rb,~Yb$^+$) \cite{Zip2010a}. Homonuclear charge transfer rates are orders of magnitudes higher \cite{Gri2009}. In our case the charge transfer is predicted to be dominantly radiative, where a photon carries away most of the 1$\,$eV of energy released in this reaction \cite{Mak2003}. The Rb$^+$ ion can be stored in the trap for hours, even without any type of cooling, which opens up  perspectives to study energetically resonant (Rb,~Rb$^+$) collisions.

 \begin{figure}
 \includegraphics[width=8.0cm] {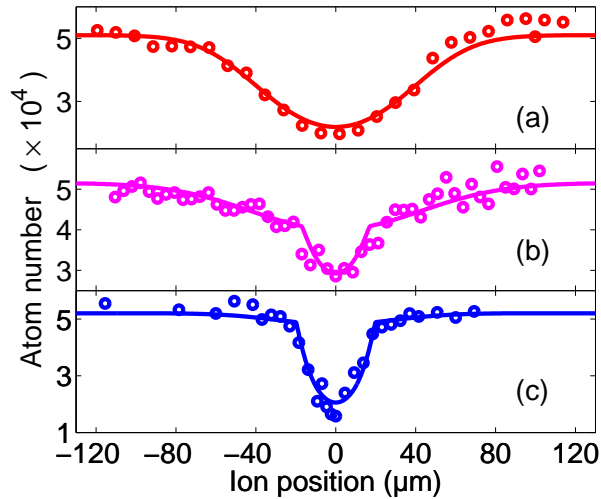}
 \caption{Number of Rb atoms remaining in the trap depending on the position of the Rb$^+$ ion relative to the center of the atom cloud. The measurement is performed with (a) a thermal cloud, (b) a partially condensed cloud and (c) an almost pure Bose-Einstein condensate. The interaction time was (a) 1.5$\,$s, (b) 1$\,$s and (c) 0.5$\,$s. The solid lines are fits, where the ion energy $E_I$ and the atom temperature $T_\textrm{Rb}$ are used as free fit parameters.}
 \label{fig4}
\end{figure}

We now show how a single ion may be used to locally probe
the atomic density distribution. By controlling the endcap voltage
a single Rb$^+$ ion is moved stepwise across the atom cloud and at each
position $x$ the atom loss is measured for a given interaction time
(Fig.~5). The measurement is performed with various Rb clouds with different condensate fractions.
We can theoretically reproduce the data shown in Fig.~5 with our model, which demonstrates our quantitative understanding of the dynamics. For this we write the total atom loss of a partly condensed cloud as a sum of the individual losses from the BEC and the thermal cloud as discussed before.
For numerical calculations we choose $\sigma_\textrm{el} = 3\times 10^{-14}\,$m$^2$
and make use of the well-known density profiles for a thermal and
a condensed atom cloud confined in a harmonic trap. The thermal
component and BEC are in equilibrium and obey $N_c /N = 1 - (T_\textrm{Rb}/
T_c)^3 $ where $T_c$ is the critical temperature and $N_c/N$ is
the condensate fraction which changes with time. The temperature $T_\textrm{Rb}$
of the atomic sample is constant, as determined by the optical
trap depth. It is used as a free fit parameter in our model together with the ion energy $E_I$, which we keep fixed for all three measurements. Fig.~5 depicts the total number of remaining particles $N(x)$ as a function of the ion's position $x$. The values for $T_\textrm{Rb}$ obtained from the fit (a) 50$\,$nK, (b) 35$\,$nK and (c) 25$\,$nK are in nice agreement with the temperatures determined separately in time-of-flight measurements. Moreover the fit suggests $E_I \approx k_B\times 14\,$mK, which is in the same range as the temperatures found in the Ba$^+$ experiments. The ion probe features a spatial resolution on the \textmu m scale and has advantages compared to absorption imaging which integrates over the line of sight.

In conclusion we have immersed cold trapped ions in a sea of
ultracold neutral atoms in a novel hybrid apparatus with high
spatial and temporal control. We have investigated the dynamics of
the atom-ion system and have extracted first elastic and inelastic
collision properties. In our present setup the collision energies
are determined by the ionic excess micromotion, which we plan to
minimize for future experiments in the ultracold regime.

The authors would like to thank Albert Frisch and Sascha Hoinka for their help during the early stage of the experiment and Thomas Busch, Tommaso Calarco, Robin Côté, Bretislav Friedrich, Bo Gao, Zbigniew Idziaszek, Tobias Sch\"atz, Wolfgang Schnitzler, and  Jaques Tempère for helpful discussions and support. We are grateful to Rudi Grimm for generous support and to Michael Drewsen and the group of Rainer Blatt for advice on the design of the ion trap. This work was supported by the Austrian Science Fund (FWF). S.S. acknowledges support from the Austrian Academy of Sciences within the DOC doctorial research fellowship program.

\bibliographystyle{apsprl}

\begin{thebibliography}{39}
\expandafter\ifx\csname natexlab\endcsname\relax\def\natexlab#1{#1}\fi

\expandafter\ifx\csname bibnamefont\endcsname\relax
 \def\bibnamefont#1{#1}\fi
\expandafter\ifx\csname bibfnamefont\endcsname\relax
  \def\bibfnamefont#1{#1}\fi
\expandafter\ifx\csname citenamefont\endcsname\relax
  \def\citenamefont#1{#1}\fi
\expandafter\ifx\csname url\endcsname\relax
  \def\url#1{\texttt{#1}}\fi
\expandafter\ifx\csname urlprefix\endcsname\relax\def\urlprefix{URL }\fi
\providecommand{\bibinfo}[2]{#2}
\providecommand{\eprint}[2][]{\url{#2}}

\bibitem[{\citenamefont{Côté}(2000)\citenamefont{Côté}}]{Cot2000a}
  \bibinfo{author}{\bibfnamefont{R.}~\bibnamefont{Côté}}, \bibinfo{journal}{Phys. Rev. Lett.}
  \textbf{\bibinfo{volume}{85}}, \bibinfo{pages}{5316} (\bibinfo{year}{2000}).

\bibitem[{\citenamefont{Massignan et~al.}(2005)\citenamefont{Massignan, Pethick, Smith}}]{Mas2005}
  \bibinfo{author}{\bibfnamefont{P.}~\bibnamefont{Massignan}},
  \bibinfo{author}{\bibfnamefont{C.~J.}~\bibnamefont{Pethick}}, \bibnamefont{and}
  \bibinfo{author}{\bibfnamefont{H.}~\bibnamefont{Smith}},  \bibinfo{journal}{Phys. Rev. A}
  \textbf{\bibinfo{volume}{71}}, \bibinfo{pages}{023606} (\bibinfo{year}{2005}).

\bibitem[{\citenamefont{Kalas et~al.}(2006)\citenamefont{Kalas, Blume}}]{Kal2006}
  \bibinfo{author}{\bibfnamefont{R.~M.}~\bibnamefont{Kalas}}, \bibnamefont{and}
  \bibinfo{author}{\bibfnamefont{D.}~\bibnamefont{Blume}},  \bibinfo{journal}{Phys. Rev. A}
  \textbf{\bibinfo{volume}{73}}, \bibinfo{pages}{043608} (\bibinfo{year}{2006}).

\bibitem[{\citenamefont{Cucchietti et~al.}(2006)\citenamefont{Cucchietti, Timmermans}}]{Cuc2006}
  \bibinfo{author}{\bibfnamefont{F.~M.}~\bibnamefont{Cucchietti}}, \bibnamefont{and}
  \bibinfo{author}{\bibfnamefont{E.}~\bibnamefont{Timmermans}},  \bibinfo{journal}{Phys. Rev. Lett.}
  \textbf{\bibinfo{volume}{96}}, \bibinfo{pages}{210401} (\bibinfo{year}{2006}).

\bibitem[{\citenamefont{Côté et~al.}(2002)\citenamefont{Côté, Kharchenko, Lukin}}]{Cot2002}
  \bibinfo{author}{\bibfnamefont{R.}~\bibnamefont{Côté}},
  \bibinfo{author}{\bibfnamefont{V.}~\bibnamefont{Kharchenko}}, \bibnamefont{and}
  \bibinfo{author}{\bibfnamefont{M.~D.}~\bibnamefont{Lukin}}, \bibinfo{journal}{Phys. Rev. Lett.}
  \textbf{\bibinfo{volume}{89}}, \bibinfo{pages}{093001} (\bibinfo{year}{2002}).

\bibitem[{\citenamefont{Staanum et~al.}(2010)\citenamefont{Staanum, Hojbjerre, Skyt, Hansen, Drewsen}}]{Sta2010}
  \bibinfo{author}{\bibfnamefont{P.~F.}~\bibnamefont{Staanum}},
  \bibinfo{author}{\bibfnamefont{K.}~\bibnamefont{Hojbjerre}},
  \bibinfo{author}{\bibfnamefont{P.}~\bibnamefont{Skyt}},
  \bibinfo{author}{\bibfnamefont{A.}~\bibnamefont{Hansen}}, \bibnamefont{and}
  \bibinfo{author}{\bibfnamefont{M.}~\bibnamefont{Drewsen}}, \bibinfo{journal}{Nat. Phys.}
  \textbf{\bibinfo{volume}{6}}, \bibinfo{pages}{271} (\bibinfo{year}{2010}).

\bibitem[{\citenamefont{Schneider et~al.}(2010)\citenamefont{Schneider, Roth, Duncker, Ernsting, Schiller}}]{Sch2010a}
  \bibinfo{author}{\bibfnamefont{T.}~\bibnamefont{Schneider}},
  \bibinfo{author}{\bibfnamefont{B.}~\bibnamefont{Roth}},
  \bibinfo{author}{\bibfnamefont{H.}~\bibnamefont{Duncker}},
  \bibinfo{author}{\bibfnamefont{I.}~\bibnamefont{Ernsting}}, \bibnamefont{and}
  \bibinfo{author}{\bibfnamefont{S.}~\bibnamefont{Schiller}}, \bibinfo{journal}{Nat. Phys.}
  \textbf{\bibinfo{volume}{6}}, \bibinfo{pages}{275} (\bibinfo{year}{2010}).

\bibitem[{\citenamefont{Chin et~al.}(2010)\citenamefont{Chin, Grimm, Julienne, Tiesinga}}]{Chi2010}
  \bibinfo{author}{\bibfnamefont{C.}~\bibnamefont{Chin}},
  \bibinfo{author}{\bibfnamefont{R.}~\bibnamefont{Grimm}},
  \bibinfo{author}{\bibfnamefont{P.~S.}~\bibnamefont{Julienne}}, \bibnamefont{and}
  \bibinfo{author}{\bibfnamefont{E.}~\bibnamefont{Tiesinga}}, \bibinfo{journal}{Rev. Mod. Phys.}
  \textbf{\bibinfo{volume}{82}}, \bibinfo{pages}{1225} (\bibinfo{year}{2010}).

\bibitem[{\citenamefont{Idziaszek et~al.}(2009)\citenamefont{Idziaszek, Calarco, Julienne, Simoni}}]{Idz2009}
  \bibinfo{author}{\bibfnamefont{Z.}~\bibnamefont{Idziaszek}},
  \bibinfo{author}{\bibfnamefont{T.}~\bibnamefont{Calarco}},
  \bibinfo{author}{\bibfnamefont{P.~S.}~\bibnamefont{Julienne}}, \bibnamefont{and}
  \bibinfo{author}{\bibfnamefont{A.}~\bibnamefont{Simoni}},  \bibinfo{journal}{Phys. Rev. A}
  \textbf{\bibinfo{volume}{79}}, \bibinfo{pages}{010702} (\bibinfo{year}{2009}).

\bibitem[{\citenamefont{Cetina et~al.}(2007)\citenamefont{Cetina, Grier, Campbell, Chuang, Vuletic}}]{Cet2007}
  \bibinfo{author}{\bibfnamefont{M.}~\bibnamefont{Cetina}},
  \bibinfo{author}{\bibfnamefont{A.}~\bibnamefont{Grier}},
  \bibinfo{author}{\bibfnamefont{J.}~\bibnamefont{Campbell}},
  \bibinfo{author}{\bibfnamefont{I.}~\bibnamefont{Chuang}},  \bibnamefont{and}
  \bibinfo{author}{\bibfnamefont{V.}~\bibnamefont{Vuletic}}, \bibinfo{journal}{Phys. Rev. A}
  \textbf{\bibinfo{volume}{76}}, \bibinfo{pages}{041401} (\bibinfo{year}{2007}).

\bibitem[{\citenamefont{Grier et~al.}(2009)\citenamefont{Grier, Cetina, Orucevic, Vuletic}}]{Gri2009}
  \bibinfo{author}{\bibfnamefont{A.~T.}~\bibnamefont{Grier}},
  \bibinfo{author}{\bibfnamefont{M.}~\bibnamefont{Cetina}},
  \bibinfo{author}{\bibfnamefont{F.}~\bibnamefont{Orucevic}}, \bibnamefont{and}
  \bibinfo{author}{\bibfnamefont{V.}~\bibnamefont{Vuletic}}, \bibinfo{journal}{Phys. Rev. Lett.}
  \textbf{\bibinfo{volume}{102}}, \bibinfo{pages}{223201} (\bibinfo{year}{2009}).

\bibitem[{\citenamefont{Zipkes et~al.}(2010)\citenamefont{Zipkes, Palzer, Sias, K\"{o}hl}}]{Zip2010}
  \bibinfo{author}{\bibfnamefont{C.}~\bibnamefont{Zipkes}},
  \bibinfo{author}{\bibfnamefont{S.}~\bibnamefont{Palzer}},
  \bibinfo{author}{\bibfnamefont{C.}~\bibnamefont{Sias}}, \bibnamefont{and}
  \bibinfo{author}{\bibfnamefont{M.}~\bibnamefont{K\"{o}hl}},
  \bibinfo{journal}{Nature} \textbf{\bibinfo{volume}{464}}, \bibinfo{pages}{388-391}
  (\bibinfo{year}{2010}).

\bibitem[{\citenamefont{Zipkes et~al.}(2010)\citenamefont{Zipkes, Palzer, Ratschbacher, Sias, K\"{o}hl}}]{Zip2010a}
  \bibinfo{author}{\bibfnamefont{C.}~\bibnamefont{Zipkes}},
  \bibinfo{author}{\bibfnamefont{S.}~\bibnamefont{Palzer}},
  \bibinfo{author}{\bibfnamefont{L.}~\bibnamefont{Ratschbacher}},
  \bibinfo{author}{\bibfnamefont{C.}~\bibnamefont{Sias}}, \bibnamefont{and}
  \bibinfo{author}{\bibfnamefont{M.}~\bibnamefont{K\"{o}hl}},
  \bibinfo{journal}{arXiv:1005.3846v1}
  (\bibinfo{year}{2010}).


\bibitem[{\citenamefont{Teachout et~al.}(1971)\citenamefont{Teachout, Pack}}]{Tea1971}
  \bibinfo{author}{\bibfnamefont{R.}~\bibnamefont{Teachout}}, \bibnamefont{and}
  \bibinfo{author}{\bibfnamefont{R.}~\bibnamefont{Pack}},  \bibinfo{journal}{Atomic Data}
  \textbf{\bibinfo{volume}{3}}, \bibinfo{pages}{195} (\bibinfo{year}{1971}).

\bibitem[{\citenamefont{Côté et~al.}(2000)\citenamefont{Côté, Dalgarno}}]{Cot2000b}
  \bibinfo{author}{\bibfnamefont{R.}~\bibnamefont{Côté}}, \bibnamefont{and}
  \bibinfo{author}{\bibfnamefont{A}~\bibnamefont{Dalgarno}}, \bibinfo{journal}{Phys. Rev. A}
  \textbf{\bibinfo{volume}{62}}, \bibinfo{pages}{012709} (\bibinfo{year}{2000}).

\bibitem[{\citenamefont{Thalhammer}(2005)\citenamefont{Thalhammer}}]{Tha2005}
    \bibinfo{author}{\bibfnamefont{G.}~\bibnamefont{Thalhammer}},
    \bibinfo{author}{\bibfnamefont{M.}~\bibnamefont{Theis}},
    \bibinfo{author}{\bibfnamefont{K.}~\bibnamefont{Winkler}},
    \bibinfo{author}{\bibfnamefont{R.}~\bibnamefont{Grimm}}, \bibnamefont{and}
  \bibinfo{author}{\bibfnamefont{H.}~\bibnamefont{Hecker Denschlag}}, \bibinfo{journal}{Phys. Rev. A}
  \textbf{\bibinfo{volume}{71}}, \bibinfo{pages}{033403} (\bibinfo{year}{2005}).

\bibitem[{\citenamefont{Schmid et~al.}(2006)\citenamefont{Schmid, Thalhammer, Winkler, Lang, Hecker Denschlag}}]{Sch2006}
  \bibinfo{author}{\bibfnamefont{S.}~\bibnamefont{Schmid}},
  \bibinfo{author}{\bibfnamefont{G.}~\bibnamefont{Thalhammer}},
  \bibinfo{author}{\bibfnamefont{K.}~\bibnamefont{Winkler}},
  \bibinfo{author}{\bibfnamefont{F.}~\bibnamefont{Lang}},  \bibnamefont{and}
  \bibinfo{author}{\bibfnamefont{J.}~\bibnamefont{Hecker Denschlag}},
  \bibinfo{journal}{New. J. Phys.} \textbf{\bibinfo{volume}{8}}, \bibinfo{pages}{159}
  (\bibinfo{year}{2006}).

\bibitem[{\citenamefont{Berkeland et~al.}(1998)\citenamefont{Berkeland, Miller, Bergquist, Itano, Wineland}}]{Ber1998}
  \bibinfo{author}{\bibfnamefont{D.~J.}~\bibnamefont{Berkeland}},
  \bibinfo{author}{\bibfnamefont{J.~D.}~\bibnamefont{Miller}},
  \bibinfo{author}{\bibfnamefont{J.~C.}~\bibnamefont{Bergquist}},
  \bibinfo{author}{\bibfnamefont{W.~M.}~\bibnamefont{Itano}}, \bibnamefont{and}
  \bibinfo{author}{\bibfnamefont{D.~J.}~\bibnamefont{Wineland}}, \bibinfo{journal}{J. App. Phys.}
  \textbf{\bibinfo{volume}{83}}, \bibinfo{pages}{5025-5033} (\bibinfo{year}{1998}).

\bibitem[{\citenamefont{Major et~al.}(2000)\citenamefont{Major, Dehmelt}}]{Maj1968}
  \bibinfo{author}{\bibfnamefont{F.~G.}~\bibnamefont{Major}}, \bibnamefont{and}
  \bibinfo{author}{\bibfnamefont{H.~G.}~\bibnamefont{Dehmelt}}, \bibinfo{journal}{Phys. Rev.}
  \textbf{\bibinfo{volume}{170}}, \bibinfo{pages}{91} (\bibinfo{year}{1968}).

\bibitem[{\citenamefont{Wineland et~al.}(1998)\citenamefont{Wineland, Monroe, Itano, Leibfried, King, Meekhof}}]{Win1998}
  \bibinfo{author}{\bibfnamefont{D.}~\bibnamefont{Wineland}},
  \bibinfo{author}{\bibfnamefont{C.}~\bibnamefont{Monroe}},
  \bibinfo{author}{\bibfnamefont{W.M.}~\bibnamefont{Itano}},
  \bibinfo{author}{\bibfnamefont{D.}~\bibnamefont{Leibfried}},
  \bibinfo{author}{\bibfnamefont{B.E.}~\bibnamefont{King}}, \bibnamefont{and}
  \bibinfo{author}{\bibfnamefont{D.M.}~\bibnamefont{Meekhof}}, \bibinfo{journal}{J. Res. Natl. Inst. Stand. Technol.}
  \textbf{\bibinfo{volume}{103}}, \bibinfo{pages}{259} (\bibinfo{year}{1998}).

\bibitem[{\citenamefont{Makarov et~al.}(2003)\citenamefont{Makarov, Côté, Michels, Smith}}]{Mak2003}
  \bibinfo{author}{\bibfnamefont{O.P.}~\bibnamefont{Makarov}},
  \bibinfo{author}{\bibfnamefont{R.}~\bibnamefont{Côté}},
  \bibinfo{author}{\bibfnamefont{H.}~\bibnamefont{Michels}}, \bibnamefont{and}
  \bibinfo{author}{\bibfnamefont{W.W.}~\bibnamefont{Smith}}, \bibinfo{journal}{Phys. Rev. A}
  \textbf{\bibinfo{volume}{67}}, \bibinfo{pages}{042705} (\bibinfo{year}{2003}).


\end{thebibliography}

\end{document}